\documentclass[prd,superscriptaddress,twocolumn,epsf,nofootinbib]{revtex4-1}

\usepackage{amsfonts,amssymb,amsmath} 
\usepackage{mathrsfs} 
\usepackage{color} 
\usepackage{graphicx}
\usepackage{dcolumn}
\usepackage{bm}
 
\newcommand{\C}[1]{{\mathcal #1}}

\newcommand{\BF}[1]{{\mathbf #1}} 

\newcommand{\Tr}{\mathop{\rm Tr}} 
 
\newcommand{\half}{\frac 12} 
\newcommand{\third}{\frac 13} 
\newcommand{\quarter}{\frac 14}

\newcommand{\Slash}[1]{{\ooalign{\hfil#1\hfil\crcr\raise.167ex\hbox{/}}}}

\begin{document} 

\title{Low energy implications of cosmological data in $U(1)_X$ Higgs inflation} 
\author{Shinsuke Kawai} 
\email{kawai@skku.edu} 
\affiliation{Department of Physics, Sungkyunkwan University, 
Suwon 16419, Republic of Korea}
\author{Nobuchika Okada}
\email{okadan@ua.edu}
\affiliation{
Department of Physics and Astronomy, 
University of Alabama, 
Tuscaloosa, AL35487, USA} 
\author{Satomi Okada}
\email{satomi.okada@ua.edu}
\affiliation{
Department of Physics and Astronomy, 
University of Alabama, 
Tuscaloosa, AL35487, USA} 
\date{\today}
 
\begin{abstract}
A scalar field having the Coleman-Weinberg type effective potential arises in various contexts of particle physics and serves as a useful framework for discussing cosmic inflation.
According to recent studies based on the Markov chain Monte Carlo analysis, the coefficients of such an effective potential are severely constrained by the cosmological data.
We investigate the impact of this observation on the physics beyond the Standard Model, focusing on an inflationary model based on the $U(1)_X$-extended Standard Model as a well-motivated example.
We examine the parameter region that is not excluded by the Large Hadron Collider (LHC) Run-2 at 139 fb${}^{-1}$ integrated luminosity, and show that the model parameters can be further constrained by the High-Luminosity LHC experiments in the near future.
We also comment on the possible reheating mechanism and the dark matter candidates of this scenario.
\end{abstract}
 
\maketitle 
\section{Introduction}

Cosmic inflation was originally proposed in the context of grand unified theories  (GUTs) \cite{Guth:1980zm,Sato:1980yn,Kazanas:1980tx} and gravitational effective theories \cite{Nariai:1971sv,Starobinsky:1980te}.
As the simple models based on the GUT scenario turned out to be unsuccessful, and as the quantum generation mechanism of the primordial fluctuations seeding the large scale structure of the Universe was found to be enormously successful, inflationary cosmology has become a major paradigm of modern physics in its own right, not necessarily associated with its particle physics origin.
This, of course, does not mean search of the particle physics responsible for inflation is unimportant.
With the rapidly growing data from cosmological precision measurements and collider experiments, the time may now be ripe for discussing a coherent picture of the early Universe based on particle physics.

The Coleman-Weinberg mechanism \cite{Coleman:1973jx} is a natural realization of spontaneous symmetry breaking as a consequence of radiative quantum corrections.
While the mechanism is not directly relevant within the electroweak symmetry breaking of the Standard Model, it can be, in principle, responsible for symmetry breaking in theories beyond the Standard Model \cite{Iso:2009ss,Iso:2009nw}.
It is natural to suppose that the Coleman-Weinberg mechanism may have played some role in the early Universe, as cosmic inflation may well be realized by some Higgs-like scalar field that existed in the early Universe.
Indeed, the inflationary model based on the Coleman-Weinberg effective potential,
\begin{align}\label{eqn:CWVeff}
  V_{\rm eff}=\frac{\lambda}{4}\phi^4\left(1+b\ln\left[\frac{\phi}{\mu}\right]\right),
\end{align}
where $\lambda$, $b$ are dimensionless constants, $\mu$ is the renormalization scale and $\phi$ is a scalar field (inflaton), has been widely studied, although its original model is known to be disfavored by the recent cosmic microwave background (CMB) observations by more than 2-$\sigma$ 
(e.g.~\cite{Barenboim:2013wra}).
At high energies a scalar field can nonminimally couple to the Ricci scalar.
Taking the nonminimal coupling into account, it is known that the inflationary model with the Coleman-Weinberg effective potential \eqref{eqn:CWVeff} assumed in the Jordan frame gives the spectrum of the CMB consistent with the present observations within 1-$\sigma$ \cite{Okada:2010jf}.
Moreover, it has been pointed out recently \cite{Rodrigues:2020dod} that the Markov chain Monte Carlo (MCMC) analysis constrains the coefficient $b$ to be positive definite by more than 3-$\sigma$.
Although the paper \cite{Rodrigues:2020dod} focuses on the type I and type II seesaw mechanisms and concludes that the type II mechanism is favored, the implication of this observation is profound and extends beyond the seesaw model; if this claim is substantiated by further observational data, many interesting cosmological scenarios would be severely constrained or ruled out.
In this paper we discuss the implication of nonzero $b$ in a simple particle physics model beyond the Standard Model, namely the $U(1)_X$-extended Standard Model.
It is shown, by a straightforward renormalization group (RG) analysis, that the breaking scale of the $U(1)_X$ symmetry is obtained from the cosmological input, and the model parameters are constrained by the current and future Large Hadron Collider (LHC) experiments. 
 
\section{Inflation in $U(1)_X$-extended Standard Model} 
 
As an example of a simple particle physics model having the Coleman-Weinberg effective potential \eqref{eqn:CWVeff}, we consider the $U(1)_X$-extended Standard Model \cite{Appelquist:2002mw}.
The gauge group of this model is $SU(3)_c\times SU(2)_L\times U(1)_Y\times U(1)_X$, that is, the Standard Model gauge group extended with an extra $U(1)_X$.
The nonabelian groups do not mix with other groups due to gauge invariance, but the abelian groups do mix among them; even if $U(1)_Y$ and $U(1)_X$ are diagonal at some energy scale, a mixing term is generated by quantum effects.
The $U(1)$ gauge couplings may be organized into the form of a triangular matrix \cite{delAguila:1988jz} and the covariant derivative with the gauge group connections is written
\begin{align}\label{eqn:CovDeriv}
  D_\mu=&\partial_\mu-ig_3T^\alpha G^\alpha_\mu-ig_2 T^a W_\mu^a\crcr
  &-ig_1 YB_\mu-i(\widetilde g_1 Y+g_X X)Z_\mu',
\end{align}
where $G^\alpha_\mu$, $W^a_\mu$, $B_\mu$, $Z_\mu'$ are the gauge fields, $g_3$, $g_2$, $g_1$, $g_X$ are the gauge couplings, and $T^\alpha$, $T^a$, $Y$, $X$ are the corresponding generators and charges of the $SU(3)_c$, $SU(2)_L$, $U(1)_Y$, $U(1)_X$ gauge groups.
The coupling $\widetilde g_1$ arises as mixing of the two $U(1)$'s.

 
\begin{table}[t]
\begin{tabular}{l|cccc}
  &~~~$SU(3)_c$&~~~$SU(2)_L$&~~~$U(1)_Y$~~~&~~~$U(1)_X$\\
  \hline\\
  $q_L^i$&${\BF 3}$&${\BF 2}$&$\frac 16$&$\frac 16 x_H+\third x_\Phi$\\\\
  $u_R^i$&${\BF 3}$&${\BF 1}$&$\frac 23$&$\frac 23 x_H+\third x_\Phi$\\\\
  $d_R^i$&${\BF 3}$&${\BF 1}$&$-\third$&$-\third x_H+\third x_\Phi$\\\\
  $\ell_L^i$&${\BF 1}$&${\BF 2}$&$-\half$&$-\half x_H-x_\Phi$\\\\
  $e_R^i$&${\BF 1}$&${\BF 1}$&$-1$&$-x_H-x_\Phi$\\\\
  $H$&${\BF 1}$&${\BF 2}$&$\half$&$\half x_H$\\\\
  $N_R^i$&${\BF 1}$&${\BF 1}$&$0$&$-x_\Phi$\\\\
  $\Phi$&${\BF 1}$&${\BF 1}$&$0$&$2x_\Phi$\\\\
  \hline
\end{tabular}
\caption{Representations and charges of the particle contents.
The index $i=1,2,3$ denotes the generations, and $x_H$ and $x_\Phi$ are real parameters that are not fixed by anomaly cancellation conditions.}
\label{tab:Contents}
\end{table}

Besides the Standard Model particles, three singlet fermions (right-handed neutrinos) $N_R^i$ and one complex scalar $\Phi$ are introduced; the former are necessary for the theory to be free from gauge and gravitational anomalies, and the latter is responsible for breaking the $U(1)_X$ symmetry and generating Majorana masses of the singlet fermions.
The particle contents of the $U(1)_X$-extended Standard Model are summarized in TABLE \ref{tab:Contents}.
Requiring the absence of gauge and gravitational anomalies, the $U(1)_X$ charges are determined up to two real parameters $x_H$ and $x_\Phi$.
As the overall scale of $x_\Phi$ may be absorbed into redefinition of $g_X$, we shall set $x_\Phi=1$. 
The remaining $x_H$ is a real constant parameter of the model.
The special case $x_H=0$ corresponds to the $U(1)_{\rm B-L}$-extended Standard Model, whereas $x_H= -4/5$ and $x_H=-2$ correspond to the standard and flipped $SU(5)\times U(1)_X$ that may be embedded in the $SO(10)$ GUT.
In this paper we leave $x_H$ as a free parameter.
The Yukawa terms are
\begin{align}
  {\C L}_{\rm Yukawa}=&-y_u^{ij}\overline q_L^i\widetilde H u_R^j-y_d^{ij}\overline q_L^i H d_R^j\crcr
  &-y_D^{ij}\overline\ell_L^i\widetilde H N_R^j-y_e^{ij}\overline\ell_L^i H e_R^j\crcr
  &-\half y_M^i\Phi\overline N_R^{ic}N_R^i+\text{h.c.},
\end{align}
where $\widetilde H\equiv i\sigma_2 H^*$, and we work in the basis of diagonal Majorana Yukawa matrix $y_M$.
This model is a well-motivated extension of the Standard Model, as the small nonzero neutrino masses are generated by the seesaw mechanism, the mechanism of baryogenesis via leptogenesis is readily incorporated, 
and $\Phi$ is a natural candidate of the inflaton field, as we now explain.

\begin{figure*}[t]
  \includegraphics[width=85mm]{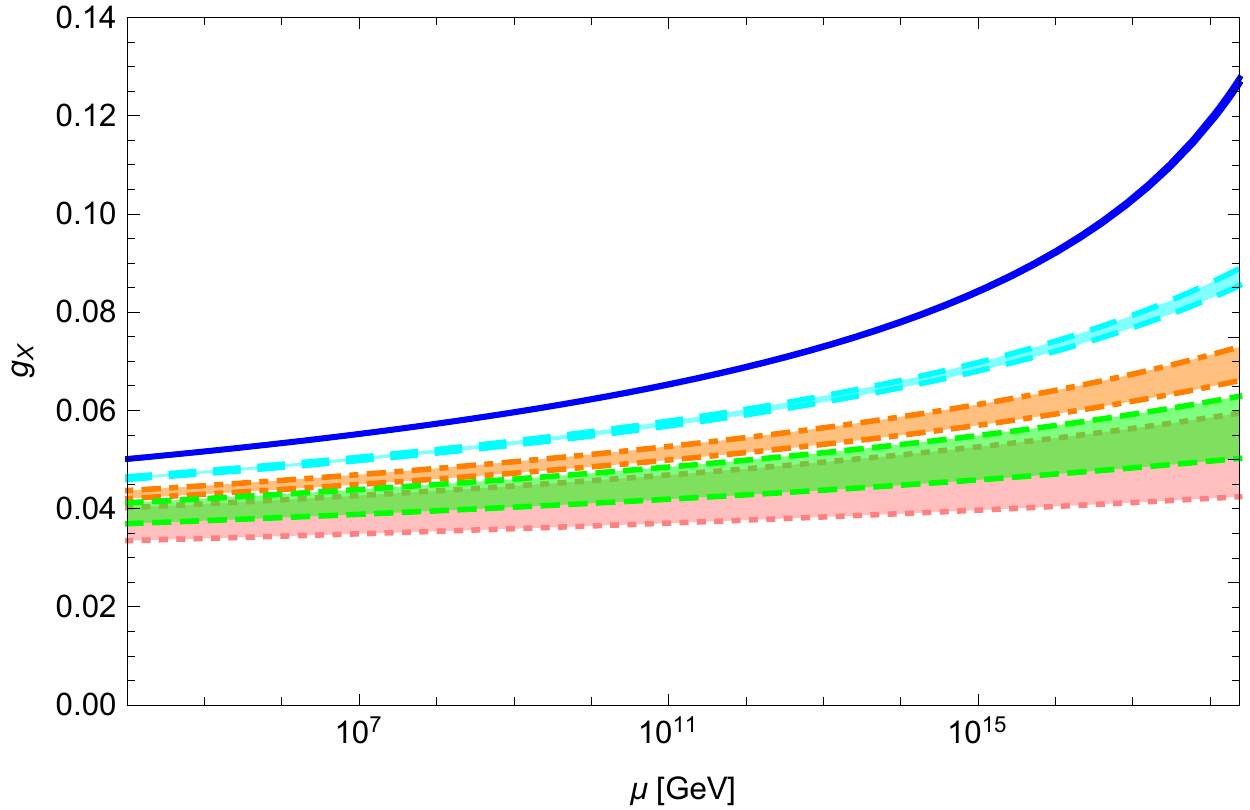}
  \includegraphics[width=90mm]{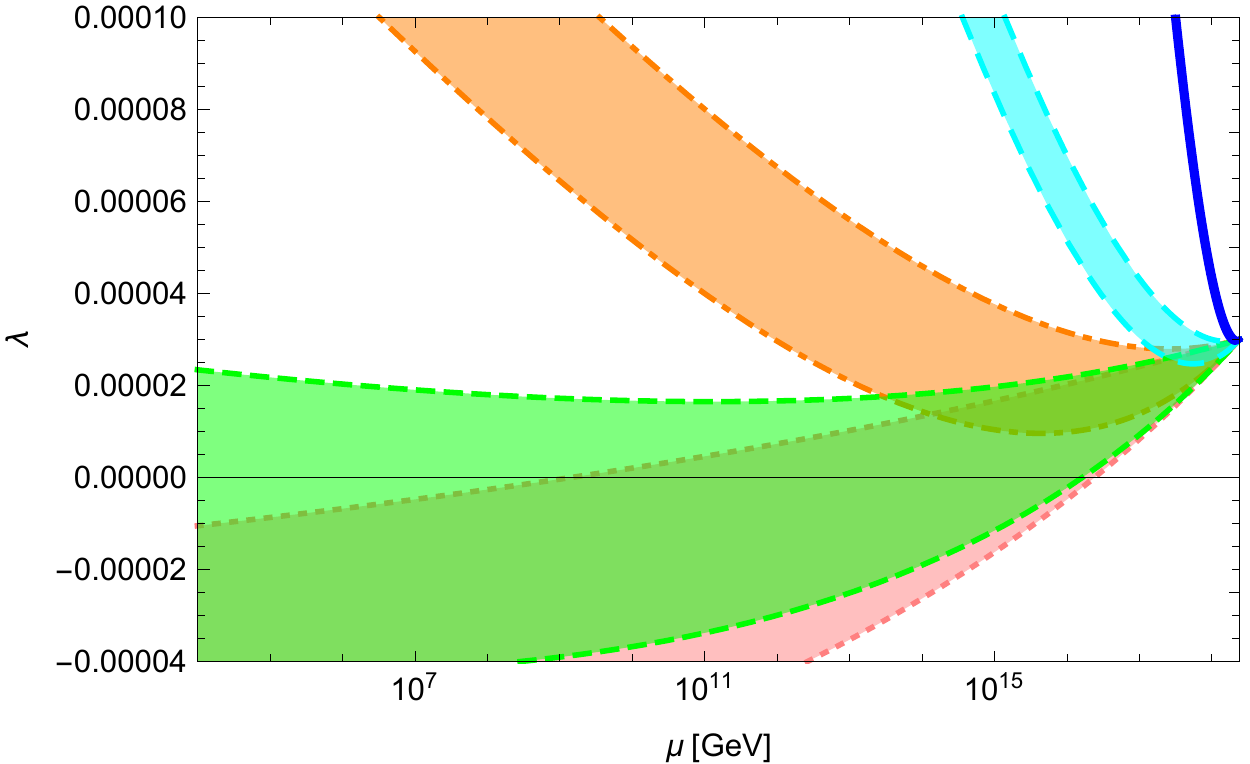}
  \caption{\label{fig:RG_flow_xH10}
The RG flows of the $U(1)_X$ gauge coupling $g_X$ (left panel) and the $U(1)_X$ Higgs self coupling $\lambda=\lambda_\Phi$ (right panel).
The value of the Majorana Yukawa coupling at $\mu=M_{\rm P}$ is varied as $y_M(M_{\rm P})=0.01$ (pink, dotted), $0.10$ (green, dashed), $0.15$ (orange, dot-dashed), $0.20$ (cyan, long-dashed) and $0.30$ (blue, solid). 
The parameter $x_H$ is fixed at 10. 
Negative $\lambda$ indicates that the $U(1)_X$ symmetry is broken by the Coleman-Weinberg mechanism.}
\end{figure*}

\begin{figure*}[t]
  \includegraphics[width=85mm]{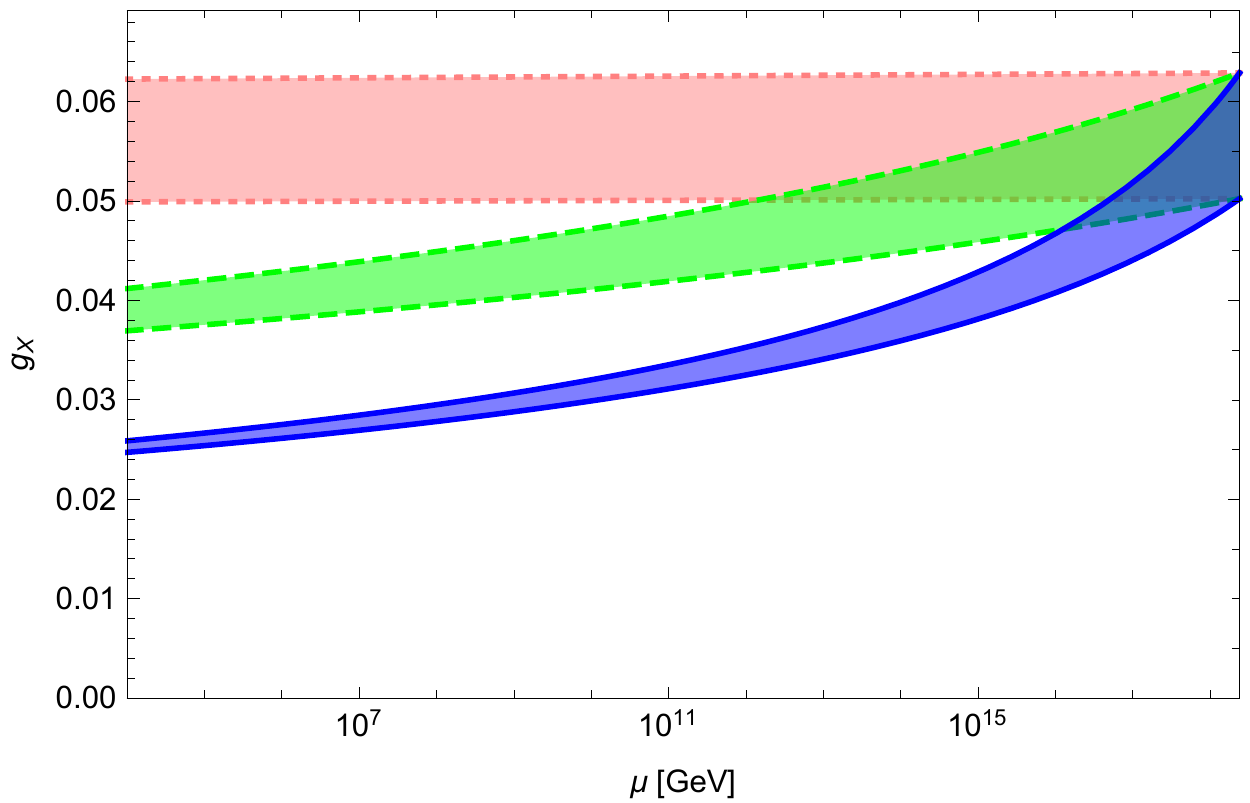}
  \includegraphics[width=90mm]{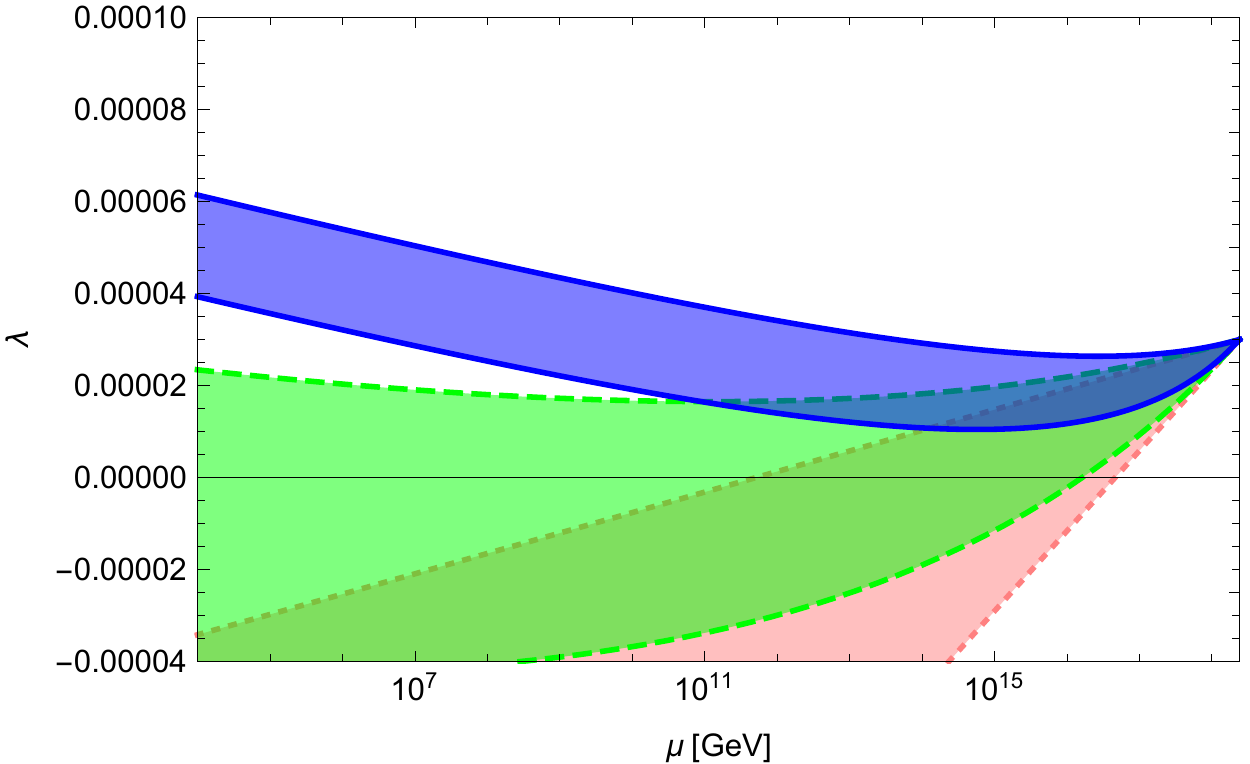}
  \caption{\label{fig:RG_flow_yM01}
The RG flows of the $U(1)_X$ gauge coupling $g_X$ (left panel) and the $U(1)_X$ Higgs self coupling $\lambda=\lambda_\Phi$ (right panel).
The values of $x_H$ is varied as $x_H=0$ (pink, dotted), 10 (green, dashed) and 20 (blue, solid). 
The Majorana Yukawa coupling at the Planck scale is fixed at $y_M(M_{\rm P})=0.1$.}
\end{figure*}

In the Higgs sector we assume the classically conformal potential
\begin{align}\label{eqn:CCPot}
  V=\lambda_\Phi(\Phi^*\Phi)^2+\lambda_H(H^\dag H)^2+\widetilde\lambda(\Phi^*\Phi)(H^\dag H),
\end{align}
where $\lambda_\Phi$ and $\lambda_H$ are the dimensionless self couplings of the singlet Higgs $\Phi$ and the doublet Higgs $H$. 
For simplicity the mixed coupling $\widetilde\lambda$ will be assumed to be small ($|\widetilde\lambda|\ll 1$) below so that the $\Phi$ and $H$ sectors may be studied separately (this can be done consistently, see e.g. \cite{Iso:2009ss}).
We will be interested in symmetry breaking by the Coleman-Weinberg mechanism.
Let us decompose $\Phi$ into the real and imaginary parts,
\begin{align}
  \Phi=\frac{1}{\sqrt 2}(\phi+i\chi)
\end{align}
and identify $\phi$ as the inflaton.
Including the 1-loop corrections, the potential for $\phi$ is
\begin{align}
  V_{\text{1-loop}}=\quarter\lambda_\Phi\phi^4+\quarter\beta_\Phi\phi^4\ln\left[\frac{\phi}{\mu}\right],
\end{align}
where $\mu$ is the renormalization scale and $\beta_\Phi$ is the 1-loop beta function given in \eqref{eqn:betaPhi}.
This is the Coleman-Weinberg effective potential \eqref{eqn:CWVeff}, with $\lambda=\lambda_\Phi$ and $b=\beta_\Phi/\lambda_\Phi$.
The action of the scalar sector pertinent to the inflationary dynamics, 
including the nonminimal coupling to the background curvature, is
\begin{align}
  S=\int d^4x\sqrt{-g}\left\{
  \half(M_{\rm P}^2+\xi\phi^2)R-\half(\partial\phi)^2-V_{\rm eff}
  \right\}.
\end{align}
Here, $\xi$ is the dimensionless coupling and $M_{\rm P}$ is the mass scale which we choose to be the reduced Planck mass $M_{\rm P}=2.24\times 10^{18}\,\text{GeV}$.
This action in the Jordan frame can be brought to the Einstein frame\footnote{In the presence of nonminimal coupling, whether the renormalization group flow is natural in the Einstein frame or in the Jordan frame is a matter of debate. We accept the latter view (e.g. \cite{George:2013iia,George:2015nza}) for concreteness here, but the distinction is insignificant in the parameter region analyzed in this paper \cite{Okada:2015lia}.} by Weyl transformation.
Then the cosmological observables can be evaluated by using the standard slow roll approximation.
The effect of positive nonminimal coupling $\xi$ is to flatten the potential in the Einstein frame, reducing the tensor amplitude relative to the scalar amplitude, thereby shifting the model prediction comfortably within the 1-$\sigma$ parameter range of the recent observational constraints \cite{Akrami:2018odb,Aghanim:2019ame}.
The MCMC analysis of \cite{Rodrigues:2020dod} shows that, for the fixed nonminimal coupling $\xi=100$, the Planck 2018 \cite{Akrami:2018odb,Aghanim:2019ame} and BICEP2-Keck Array experiments \cite{Array:2016afx,Ade:2018gkx} constrain the effective potential at the renormalization scale $\mu=M_{\rm P}$ as
\begin{align}
  \label{eqn:lambdaMP}
  &\lambda=3\times 10^{-5},\\
  \label{eqn:constraints}
  &4.92\times 10^{-7}\leq \frac{\lambda b}{4}\leq 1.90\times 10^{-6}\quad\text{(68\% C.L.)}.
\end{align}
In gauge-extended Standard Models, these constraints impose conditions on the beta function $\beta_\Phi$ at the energy scale of inflation, which then govern the physics at lower energy scales via the RG flow.

\begin{figure*}[t]
  \includegraphics[width=85mm]{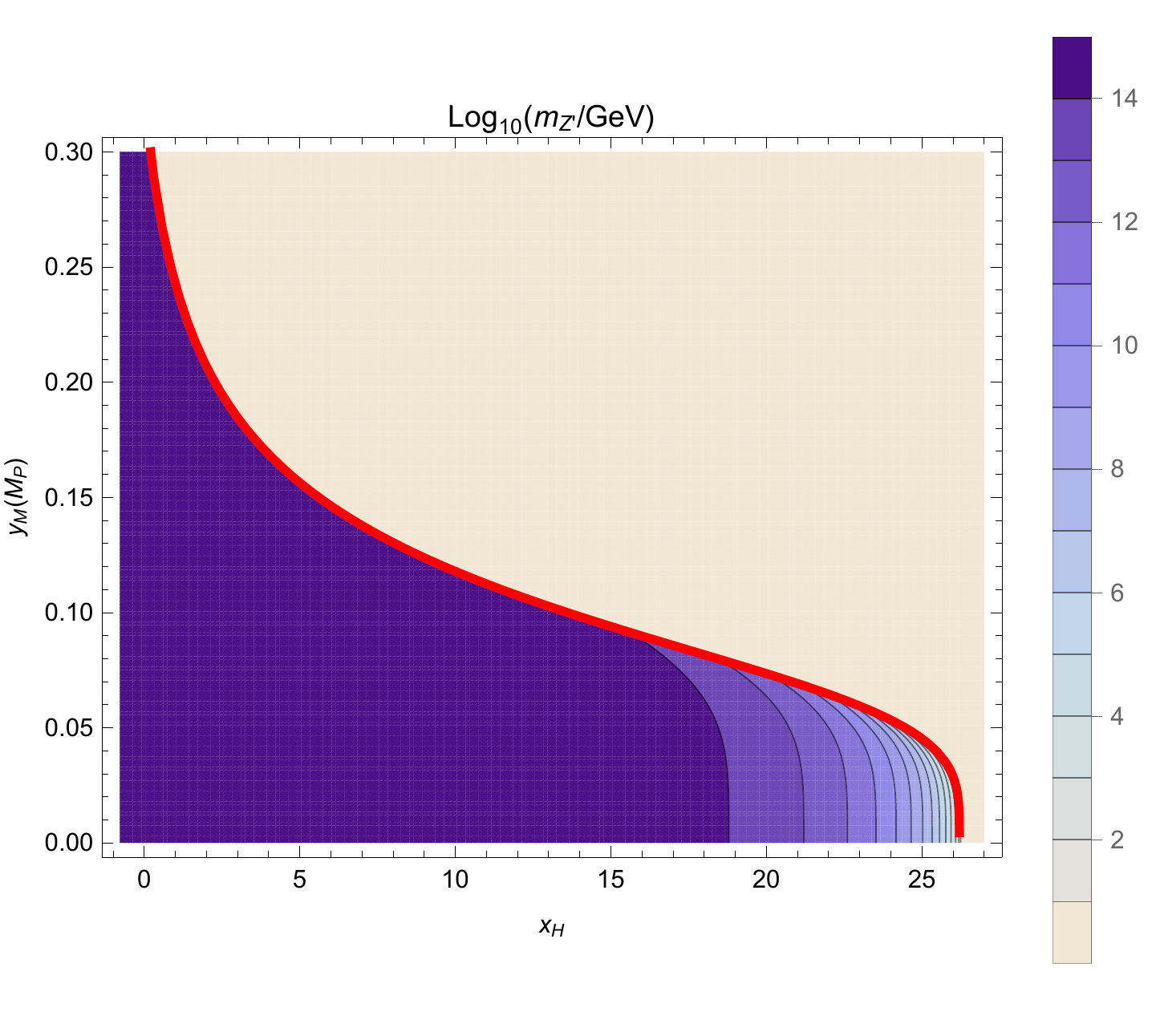}
  \includegraphics[width=85mm]{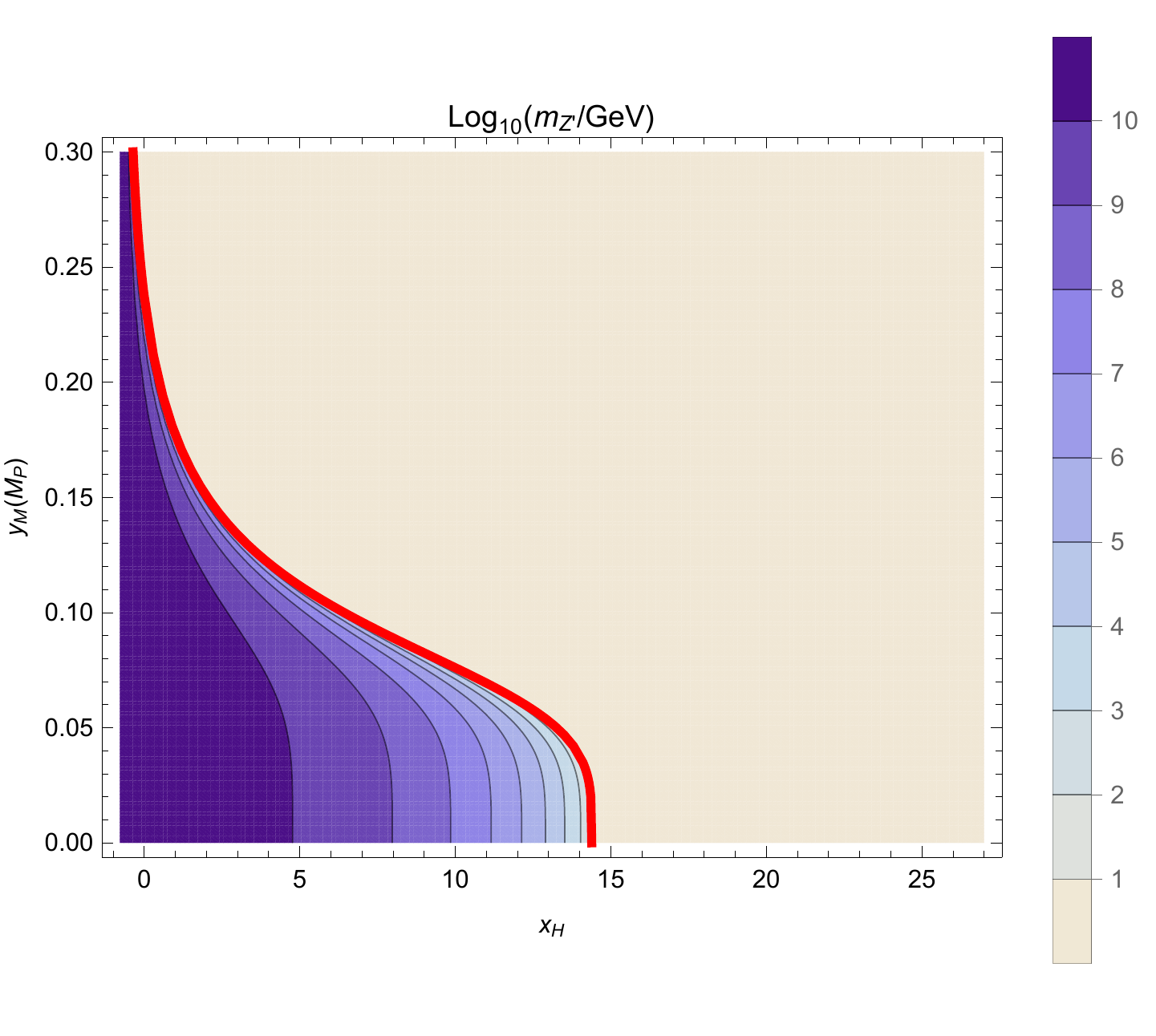}
  \caption{\label{fig:MZxContour}
The contour plot of the $Z'$ boson mass $m_{z'}$ on the $x_H$-$y_M(M_{\rm P})$ plane.
The left (right) panel uses the 1-$\sigma$ upper (lower) bound \eqref{eqn:constraints} on the coefficient $b$ of the effective potential
obtained from the cosmological MCMC analysis \cite{Rodrigues:2020dod}.
The red curve is the threshold above which the $U(1)_X$ symmetry breaking by the Coleman-Weinberg mechanism does not occur.}
\end{figure*}

An immediate consequence of identifying $\phi$ as the $U(1)_X$ Higgs field is that the conditions \eqref{eqn:lambdaMP} and \eqref{eqn:constraints} control the breaking scale of the $U(1)_X$ gauge symmetry.
Neglecting the running of $\lambda$ and $b$ for simplicity, we find from the stationarity condition $dV_{\rm eff}/d\phi=0$ that the potential \eqref{eqn:CWVeff} takes a minimum at
\begin{align}\label{eqn:phivev}
  \phi_{\rm min}=v_{\phi}=M_{\rm P}e^{-\left(\frac 1b +\frac 14\right)}.
\end{align}
The conditions \eqref{eqn:lambdaMP}, \eqref{eqn:constraints} then indicate that the $U(1)_X$ symmetry breaking by the Coleman-Weinberg mechanism takes place in the range
\begin{align}
  4.5\times 10^{11}\,\text{GeV}\leq v_\phi\leq 3.7\times 10^{16}\,\text{GeV}.
\end{align}
The right-handed neutrino masses are given by $y_M^i v_\phi/\sqrt 2$.
Thus, interestingly, this rough estimate shows that the cosmological data fitting leads to the natural seesaw scale expected from the seesaw mechanism for not too small Majorana Yukawa couplings $y_M^i$.

In the next section we present a refined analysis, including the RG flow of parameters.

\section{Implications on collider physics}

In order to simplify the analysis we neglect the mixed Higgs coupling $\widetilde\lambda$.
This can be done safely as the corresponding beta function \eqref{eqn:betamix} stays negligibly small along the RG flow.
Up to 1-loop, only the $U(1)_X$ Higgs self coupling $\lambda=\lambda_\Phi$, the $U(1)_X$ gauge coupling $g_X$ and the Majorana Yukawa coupling $y_M^i$ concern the dynamics of inflation.
We also assume for simplicity that the three diagonal components of the Majorana Yukawa coupling are degenerate, $y_M^1=y_M^2=y_M^3=y_M$.
The RG equations are then (see Appendix)
\begin{align}\label{eqn:RGElambda}
  &\frac{d\lambda}{d\ln\phi}\equiv\,\beta_\Phi\cr 
  &=\frac{1}{16\pi^2}\Big(20\lambda^2-3 y_M^4
  +96 g_X^4+6\lambda \left(y_M^2-8g_X^2\right)\Big),\\
\label{eqn:RGEgX}
  &\frac{dg_X}{d\ln\phi}=\frac{g_X^3}{16\pi^2}\left(12+\frac{32}{3}x_H+\frac{41}{6}x_H^2\right),\\
\label{eqn:RGEyM}
  &\frac{dy_M}{d\ln\phi}=\frac{y_M}{16\pi^2}\left(\frac 52 y_M^2-6g_X^2\right).
\end{align}
The boundary conditions are set by the cosmological data fitting at the scale of inflation $\mu=M_{\rm P}$.
The value of $\lambda$ is fixed by \eqref{eqn:lambdaMP}, and the condition \eqref{eqn:constraints} constrains $\beta_\Phi$, which relates the values of $y_M$ and $g_X$ at $\mu=M_{\rm P}$.
We shall regard the value of $y_M$ at $\mu=M_{\rm P}$ as an input parameter and $g_X$ to be determined within the likelihood of 68\% confidence level.
The condition \eqref{eqn:lambdaMP} suggests $\lambda\ll g_X^2, y_M^2$.
When $96 g_X^2\gg 3 y_M^4$, from the beta function \eqref{eqn:RGElambda} one can see that the constraints \eqref{eqn:constraints} give conditions on the gauge coupling at $\mu=M_{\rm P}$,
\begin{align}
  0.0424\leq g_X\leq 0.0595,
\end{align}
with the center value $g_X=0.0509$.
The RG-improved effective potential is
\begin{align}\label{eqn:quarticV}
  V(\phi)=\quarter\lambda(\phi)\phi^4
\end{align}
and a potential minimum is characterized by the stationarity condition $d V(\phi)/d\phi=0$, which gives
\begin{align}\label{eqn:stationarity}
  \beta_\Phi+4\lambda=0.
\end{align}
If the $U(1)_X$ symmetry is broken by the Coleman-Weinberg mechanism, $\lambda$ becomes negative at the potential minimum.
The breaking scale is the vacuum expectation value $v_\phi$ of the $U(1)_X$ Higgs field $\phi$ at the minimum, which is found by solving the condition \eqref{eqn:stationarity} for $\phi$.
The inflaton ($U(1)_X$ Higgs boson) mass 
$m_\phi^2=d^2V(\phi)/d\phi^2|_{\phi=v_\phi}$ and the $Z'$ boson mass 
$M_{Z'}=2v_\phi g_X$ at the breaking scale are found by solving the RG equation from $\mu=M_{\rm P}$ down to $\mu=v_\phi$.

\subsection{RG flow and the Coleman-Weinberg mechanism}

\begin{figure*}[t]
  \includegraphics[width=85mm]{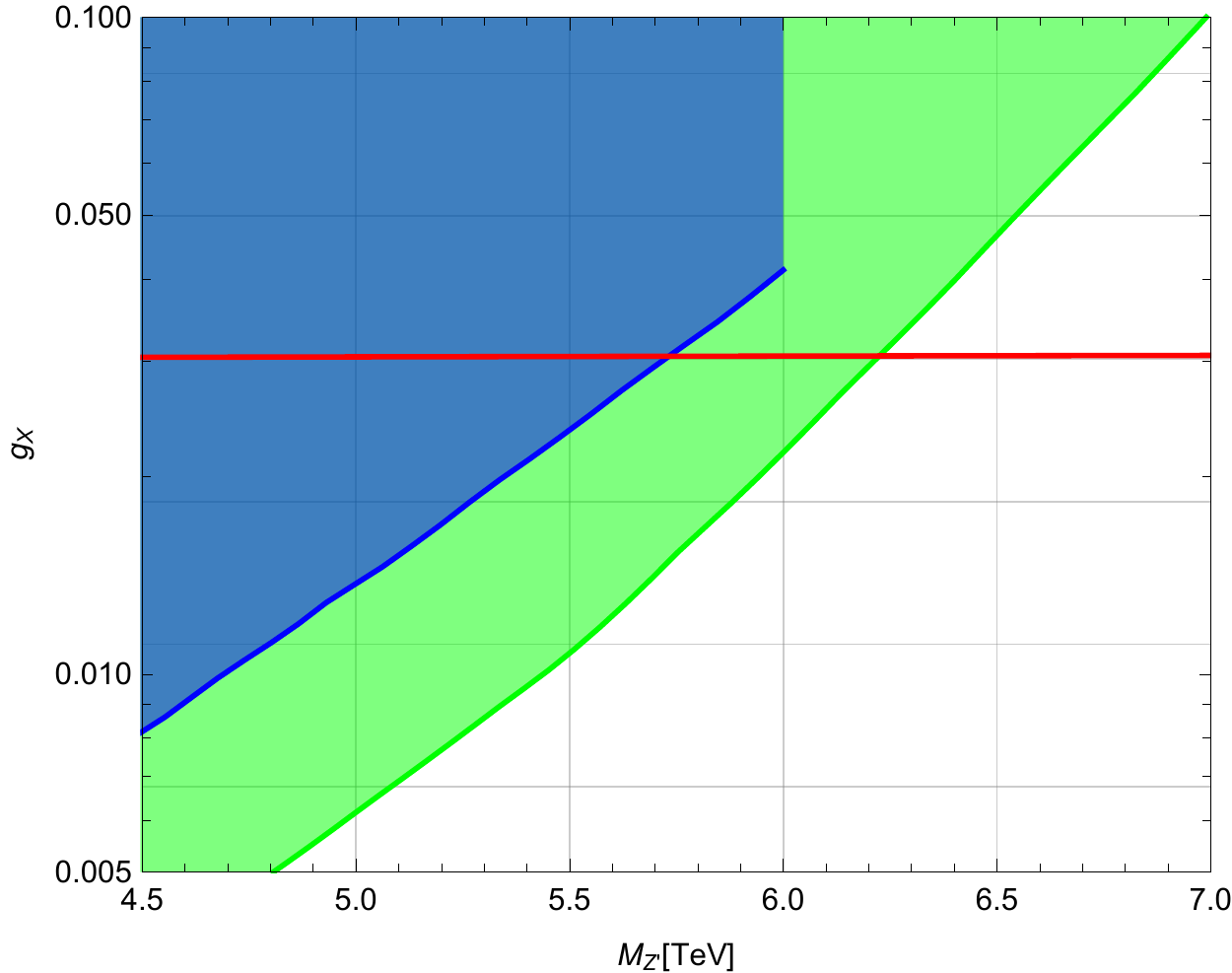}
   \includegraphics[width=85mm]{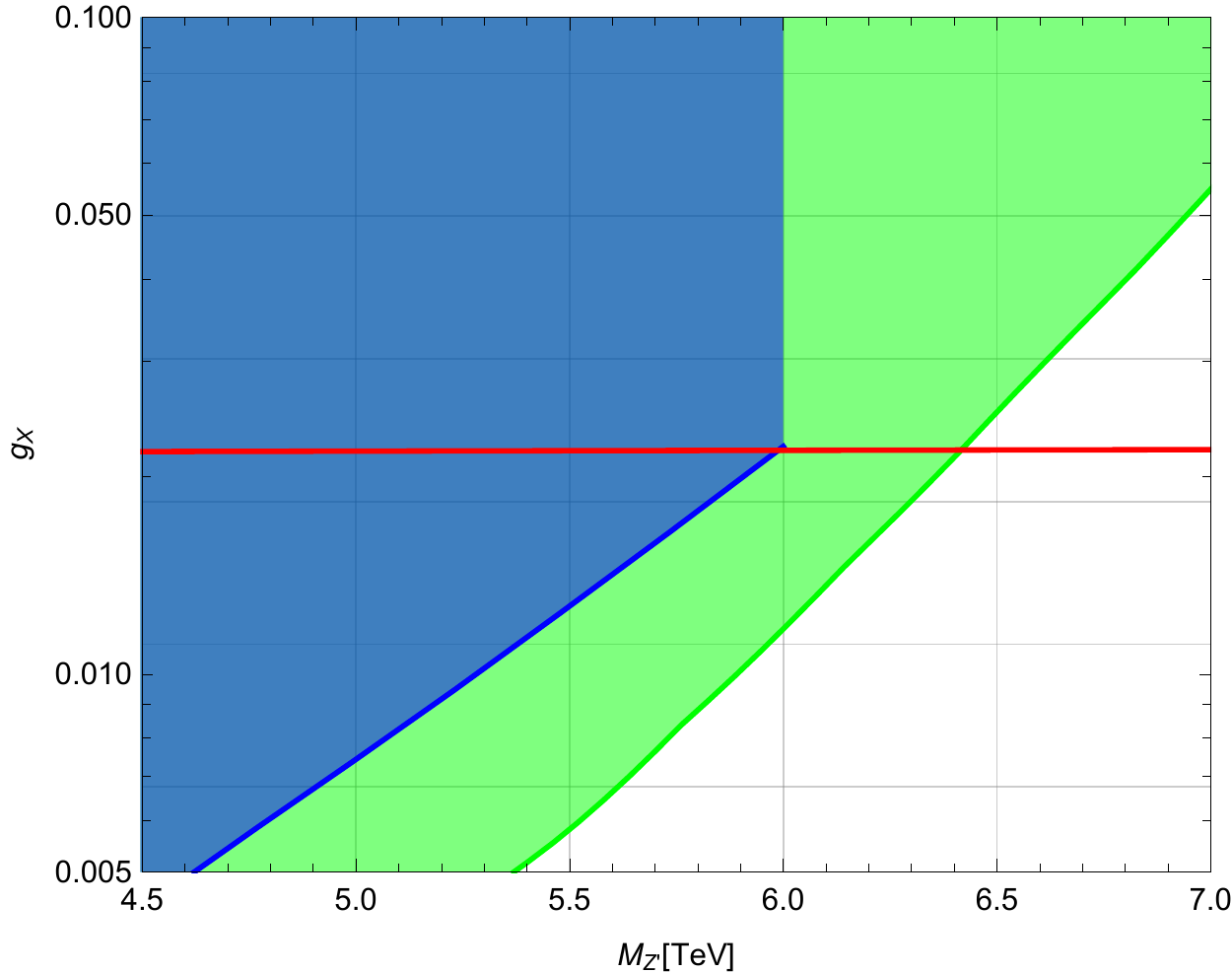}
  \caption{\label{fig:LHC_Bounds}
Constraints on the $Z'$ boson mass and the $U(1)_X$ gauge coupling by the LHC Run-2 (blue), and the region which will be searched by the HL-LHC (green). 
The left panel uses the 1-$\sigma$ lower bound and the right panel uses the 1-$\sigma$ upper bound value set by the cosmological data fitting \eqref{eqn:constraints}.
The red lines are the prediction of the inflationary model, with the Yukawa coupling at $\mu=M_{\rm P}$ chosen to be $y_M(M_{\rm P})=0.04$.
$M_{\rm Z'}=6.0\;{\rm TeV}$ is the upper bound of the $Z'$ boson mass searched by the LHC Run-2 with 139 ${\rm fb}^{-1}$ integrated luminosity.
}
\end{figure*}

The solutions to the RG equations \eqref{eqn:RGElambda}, \eqref{eqn:RGEgX}, \eqref{eqn:RGEyM}, with the boundary conditions set at $\mu=M_{\rm P}$, are shown in FIG.~1 and FIG.~2.
At the boundary $\lambda$ is fixed by \eqref{eqn:lambdaMP} and the gauge coupling $g_X$ is given by the 1-$\sigma$ constraints \eqref{eqn:constraints} through the beta function \eqref{eqn:RGElambda}.
The Yukawa coupling at $\mu=M_{\rm P}$ and the $x_H$ parameter are treated as input parameters.
FIG.~1 shows the behavior of the gauge coupling $g_X$ (the left panel) and the $U(1)_X$ Higgs self coupling $\lambda$ (the right panel), as $y_M(M_{\rm P}$) is varied as $0.01, 0.10, 0.15, 0.20$ and $0.30$, whereas $x_H$ is fixed at $x_H=10$.
For small $y_M(M_{\rm P})$, $\lambda$ is seen to become negative at lower energy, indicating that the symmetry breaking takes place by the Coleman-Weinberg mechanism.
For larger values of $y_M(M_{\rm P})$, $\lambda$ stays positive all the way and thus there is no symmetry breaking by the Coleman-Weinberg mechanism.
This implies that instead of the classically conformal Higgs potential \eqref{eqn:CCPot} one needs to consider the renormalizable potential implementing the Higgs mechanism,
\begin{align}
  V=&\lambda_\Phi\left(\Phi^*\Phi-\half v_\Phi^2\right)^2
  +\lambda_H\left(H^\dag H-\half v_H^2\right)^2\crcr
  &+\widetilde\lambda\left(\Phi^*\Phi-\half v_\Phi^2\right)\left(H^\dag H-\half v_H^2\right),
\end{align}
where $v_H=246\;{\rm GeV}$ is the Standard Model Higgs vacuum expectation value and $v_\Phi$ is a symmetry breaking scale of $U(1)_X$;
the potential involves extra parameters and the scenario becomes less predictive.
We thus focus on the more interesting case in which the $U(1)_X$ symmetry breaking is realized by the Coleman-Weinberg mechanism.
FIG.~2 shows the behavior of $g_X$ (the left panel) and $\lambda$ (the right panel), as $y_M(M_{\rm P})$ is fixed at 0.1 and $x_H$ is varied as 0, 10, and 20.
One can see a similar tendency as in FIG.~1, that is, when $x_H$ is small the symmetry breaking takes place at higher energy scale, and as $x_H$ is increased the breaking scale $v_\phi$ becomes smaller, and then for $x_H$ larger than some critical value the Coleman-Weinberg symmetry breaking cease to occur.
Both FIG.~1 and FIG.~2 show that the $U(1)_X$ breaking scale $v_\phi$ can be as low as TeV.
It is thus interesting to discuss possible signals that may be found in colliders.

\subsection{$Z'$ boson mass}

\begin{figure*}[t]
  \includegraphics[width=85mm]{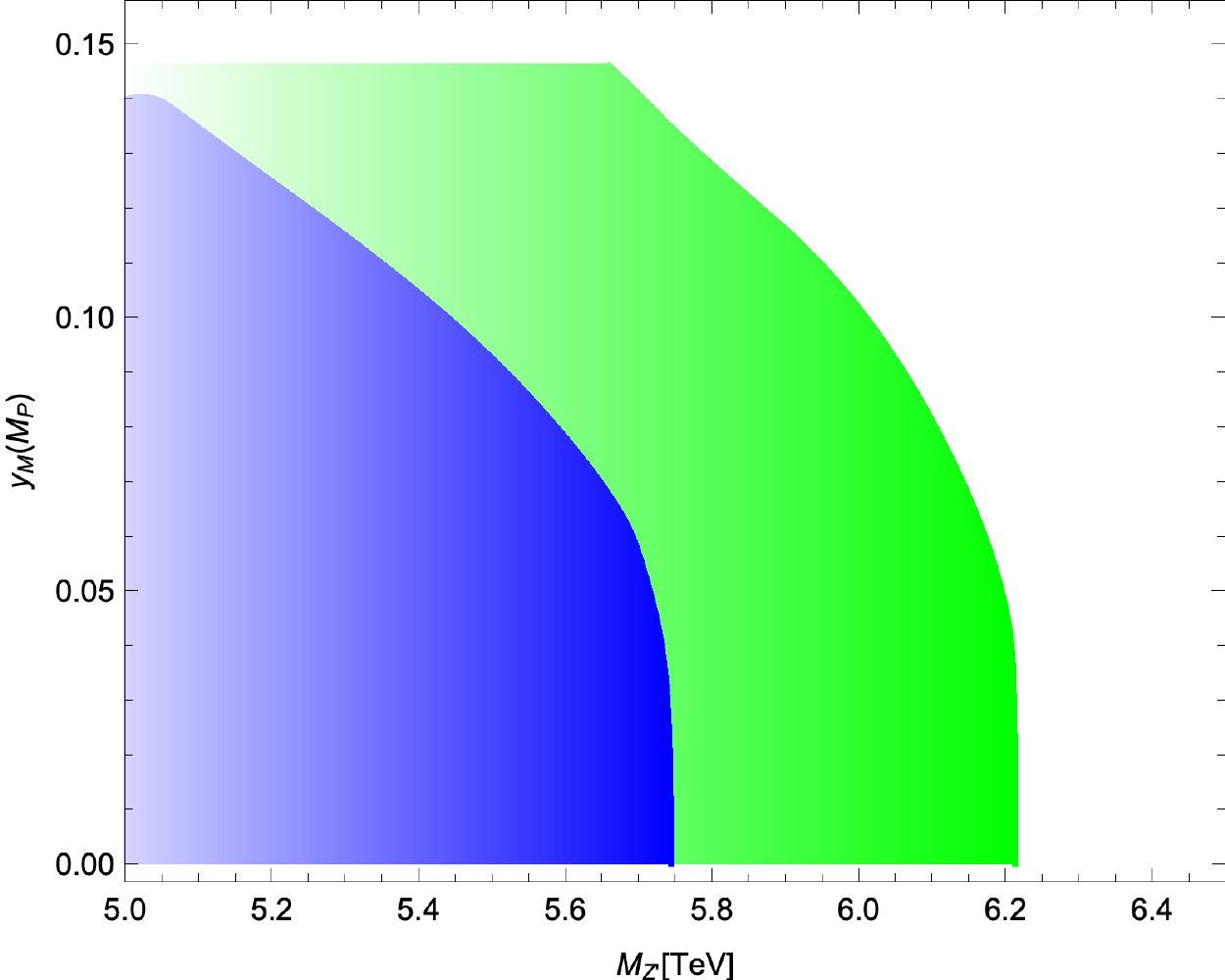}
  \includegraphics[width=85mm]{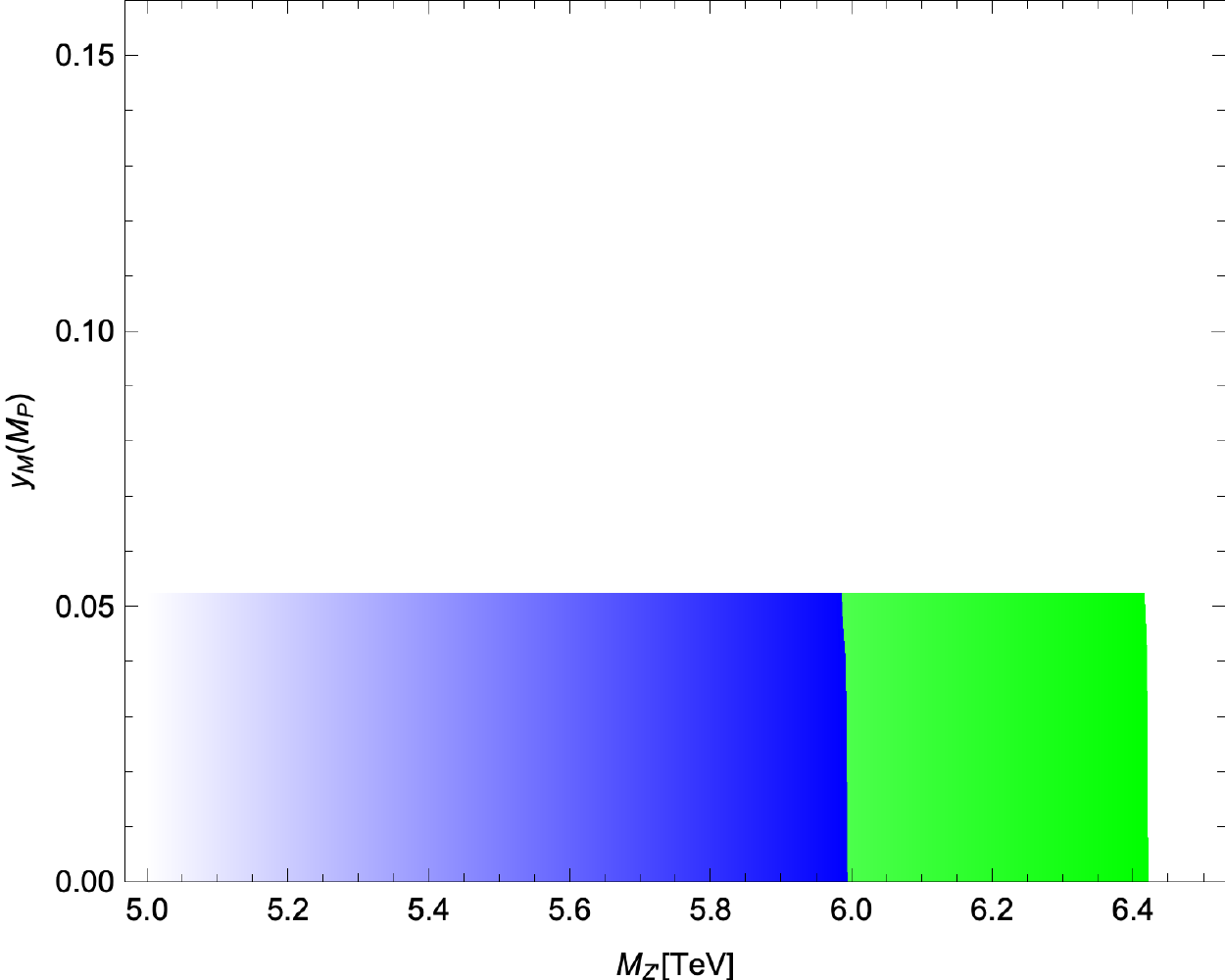}
  \caption{\label{fig:LHC_Bounds2}
Constraints on the $M_{Z'}$-$y_M(M_{\rm P})$ parameter space.
The left (right) panel uses the lower (upper) 1-$\sigma$ bound of the cosmological data \eqref{eqn:constraints}.
The blue region has been excluded by the LHC Run-2.
The green region is the expected coverage by the HL-LHC.}
\end{figure*}

When the $U(1)_X$ symmetry is broken, the $Z'$ boson becomes massive and its mass is given by $M_{Z'}=2g_Xv_\phi$.
FIG.~3 shows the $Z'$ boson mass on the $x_H$-$y_M(M_{\rm P})$ plane, computed at the $U(1)_X$ breaking scale $\phi=v_\phi$. 
The left (right) panel uses the 1-$\sigma$ lower (upper) bound of $\lambda b/4$ from the cosmological data \eqref{eqn:constraints} as the boundary condition of the renormalization group equations at $\phi=M_{\rm P}$.
As one can see from FIG.~1 and 2, the Coleman-Weinberg mechanism ceases to operate for large values of $y_M(M_{\rm P})$ and for large values of $x_H$. 
The threshold is indicated by the red curves, above which the $U(1)_X$ symmetry cannot be broken by the Coleman-Weinberg mechanism. 
The plots clearly indicate that the $Z'$ boson can be very light when $y_M(M_{\rm P})$ is small and $x_H$ is moderately large.

\subsection{Constraints by LHC}

The strongest constraints on the model parameters come from the exotic gauge boson search in hadron colliders, by the s-channel process $p+p\to Z' \to e^+e^-/\mu^+\mu^-$.
Since $g_X$ is already constrained to be small by the LHC experiments, for computing the cross section $\sigma(pp\to Z'\to e^+e^-/\mu^+\mu^-)\simeq\sigma(pp\to Z'){\rm BR}(Z'\to e^+e^-/\mu^+\mu^-)$ one may use the narrow-width approximation,
\begin{align}
  \sigma(pp\to Z')=2\sum_{q,\overline q}\int dx\int dy\, f_q(x,Q)f_{\overline q}(y,Q)\hat\sigma(\hat s),
\end{align}
with
\begin{align}
  \hat\sigma(\hat s)=\frac{4\pi^2}{3}\frac{\Gamma(Z'\to q\overline q)}{M_{Z'}}\delta(\hat s-M_{Z'}^2).
\end{align}
Here, $f_q$ and $f_{\overline q}$ are the parton distribution functions for a quark and an anti-quark, $\hat s\equiv xys$ is the invariant squared mass of the colliding quarks.
For the LHC Run-2, $\sqrt s=13\;{\rm TeV}$.
In our LHC analysis we follow \cite{Aad:2019fac}.

In FIG.~4, the red lines show the prediction $g_X$ and $M_{Z'}$ of our model at low energies, with the value of $y_M(M_{\rm P})$ fixed at $0.04$ and the values of $x_H$ varied.
The model parameters are constrained by the final result of the LHC Run-2 with 139 ${\rm fb}^{-1}$ integrated luminosity \cite{Aad:2019fac} (the blue shaded region).
The left panel uses the 1-$\sigma$ lower bound of \eqref{eqn:constraints} and the right panel uses the 1-$\sigma$ upper bound as in FIG.~3.
We also show in green color the prospect of the future $Z'$ boson search by the High-Luminosity LHC (HL-LHC) experiments\footnote{
See the ATLAS Technical Design Report \cite{ATLAS-TDR-027}.
We interpret the prospective bound as the upper bound on $g_X$ for a given value of $x_H$ \cite{Das:2019fee}.
}, with $\sqrt s=14\;{\rm TeV}$ and the goal integrated luminosity 3000 ${\rm fb}^{-1}$.
For $y_M(M_{\rm P})=0.04$ and the 1-$\sigma$ constraints \eqref{eqn:constraints}, the parameter range for $M_{Z'} < 5.7-6.0\;{\rm TeV}$ is already seen to be excluded by the LHC Run-2 experiments.
The HL-LHC covers the parameter region up to $M_{Z'} < 6.2-6.4\;{\rm TeV}$.

FIG.~5 shows on the $M_{Z'}$-$y_M(M_{\rm P})$ plane the parameter region already excluded by the LHC Run-2, indicated in blue, and the prospect of the region covered by the HL-LHC (green), as both of our model parameters $y_M(M_{\rm P})$ and $x_H$ are scanned.
The left (right) panel shows the results using the 1-$\sigma$ lower (upper) bound of \eqref{eqn:constraints}, likewise in FIG.~3 and FIG.~4.
Phenomenologically interesting parameter region is $y_M(M_{\rm P})\lesssim 0.05$ and $M_{Z'}\simeq 6\;{\rm TeV}$, which is not excluded by the LHC Run-2 experiments and in the prospect coverage of the HL-LHC.
The upper bound of $y_M(M_{\rm P})$ is due to our assumption of the Coleman-Weinberg symmetry breaking.
For larger values of $y_M(M_{\rm P})$ the symmetry breaking does not occur; see FIG.~1.

\subsection{Other aspects of the model}

Let us conclude this section by commenting on the mechanism of reheating, the scenario of baryogenesis, and dark matter candidates that are most plausible in this model.

\subsubsection{Reheating temperature}

The mass of the $Z'$ gauge boson is $M_{Z'}=2g_Xv_\phi$ and that of the right-handed neutrinos is $y_M^iv_\phi/\sqrt 2$.
The square of the inflaton mass is, using \eqref{eqn:quarticV}, 
\begin{align}
  M_\phi^2=&\frac{d^2V_{\rm eff}}{d\phi^2}\Bigg|_{\phi=v_\phi}
  =\left[\left(\beta_\Phi+\frac 14\frac{d\beta_\Phi}{d\ln\phi}\right)\phi^2\right]_{\phi=v_\phi}\crcr
  \simeq&\beta_\Phi(v_\phi)v_\phi^2.
\end{align}
As the $96g_X^4$ term dominates the beta function $\beta_\Phi$ in Eq.\eqref{eqn:RGElambda}, the inflaton mass is evaluated as $M_\phi\simeq \sqrt{3/2}g_X M_{Z'}/\pi\ll M_{Z'}$.
The inflaton cannot decay into the heavier $Z'$ boson.
Being a Standard Model gauge singlet, it cannot decay into the Standard Model gauge bosons either.
Thus the dominant decay channel of the inflaton into the Standard Model particles is through the mixing with the Standard Model Higgs field \cite{Oda:2017zul}.

The mixing arises from the last term of the classically conformal potential \eqref{eqn:CCPot} which we thus far have ignored. 
Using the unitary gauge $H=(0, h/\sqrt 2)$ the potential is written
\begin{align}
  V=\quarter\left(\lambda_\Phi(\phi)\phi^4+\lambda_H h^4+\widetilde\lambda\phi^2 h^2\right),
\end{align}
in which the running of $\lambda_\Phi(\phi)$ is indicated explicitly as it is crucial for the Coleman-Weinberg symmetry breaking. 
The quantum corrections for $\lambda_H\sim{\mathcal O}(0.1)$ and for $\widetilde\lambda$ are negligible.
From the stationarity conditions $\partial V/\partial h=0$ and $\partial V/\partial\phi=0$ we have
\begin{align}\label{eqn:stath}
  &\widetilde\lambda=\, -2\lambda_H\left(\frac{v_h}{v_\phi}\right)^2,\\
  \label{eqn:statphi}
  &\beta_\Phi+4\lambda_\Phi+2\widetilde\lambda\left(\frac{v_h}{v_\phi}\right)^2=0.
\end{align}
As a benchmark, let us consider $M_{Z'}=6\;{\rm TeV}$ and $g_X=0.03$, giving $v_\phi=10^5\;{\rm GeV}$.
The Standard Model Higgs vacuum expectation value is $v_h=\langle h\rangle=246\;{\rm GeV}\ll v_\phi$.
Then from \eqref{eqn:stath} one finds $|\widetilde\lambda|(v_h/v_\phi)^2\simeq\lambda_H(v_h/v_\phi)^4\ll\lambda_\Phi\sim 10^{-5}$.
The last term of \eqref{eqn:statphi} is thus negligibly small, justifying our treatment of the mixed coupling term in our analysis above.
Redefining the $h$ and $\phi$ fields around the expectation values as $h\to h+v_h$ and $\phi\to\phi+v_\phi$, the mass term of the potential is written
\begin{align}
  V\supset\half (h\; \phi)\begin{pmatrix}
    m_h^2 & \widetilde m^2\\
    \widetilde m^2 & m_\phi^2
  \end{pmatrix}
  \begin{pmatrix}
    h \\ \phi
  \end{pmatrix},
\end{align}
with the mass matrix elements
\begin{align}\label{eqn:mhsq}
  m_h^2=&\frac{\partial^2 V}{\partial h^2}\Bigg\vert_{h=v_h,\phi=v_\phi}=2\lambda_Hv_h^2=-\widetilde\lambda v_\phi^2,\\
  \label{eqn:mmixsq}
  \widetilde m^2=&\frac{\partial^2 V}{\partial h\partial\phi}\Bigg\vert_{h=v_h,\phi=v_\phi}=\widetilde\lambda v_hv_\phi,\\
  \label{eqn:mphisq}
  m_\phi^2\simeq &\frac{3}{2\pi^2}g_X^2M_{Z'}^2.
\end{align}
The mass matrix is diagonalized by rotating the fields,
\begin{align}
  \begin{pmatrix}
    h\\\phi
  \end{pmatrix}
  =\begin{pmatrix}
    \cos\theta & \sin\theta\\ -\sin\theta & \cos\theta
  \end{pmatrix}
  \begin{pmatrix}
    \widetilde h \\ \widetilde\phi
  \end{pmatrix}
\end{align}
with the angle given by $\tan 2\theta=2\widetilde m^2/(m_\phi^2-m_h^2)$.
Let us, from FIG.~4, choose
\begin{align}\label{eqn:gXrange}
	0.022\leq g_X\leq 0.031
\end{align}
as the benchmark values of the low energy $U(1)_X$ gauge coupling compatible with the 1-$\sigma$ cosmological data.
For $M_{Z'}=6\;{\rm GeV}$, the inflaton mass is given by \eqref{eqn:mphisq} as
\begin{align}\label{eqn:mphirange}
	51.5\;{\rm GeV}\leq m_\phi\leq 72.5\;{\rm GeV}.
\end{align}
The Standard Model Higgs mass is $m_h=125\;{\rm GeV}$.
Thus for our parameter choice the mixing angle is very small,
\begin{align}\label{eqn:thetarange}
	2.17\times 10^{-3}\leq\sin\theta\leq 3.83\times 10^{-3},
\end{align}
and consequently the field $\widetilde h$ is almost $h$, and $\widetilde\phi$ is almost $\phi$.

The `inflaton' $\widetilde\phi$ dominantly decay to $f\overline f=b\overline b, c\overline c$ and $\tau\overline\tau$ with the coupling $y_f\sin\theta/\sqrt 2$, where $y_f$ are the corresponding Standard Model Yukawa couplings.
The total decay width of $\widetilde\phi$ is
\begin{align}
	\Gamma_{\widetilde\phi}\simeq\frac{m_\phi}{8\pi}\left\{3\left(\frac{m_b}{v_h}\right)^2+3\left(\frac{m_c}{v_h}\right)^2+\left(\frac{m_\tau}{v_h}\right)^2\right\}\sin^2\theta,
\end{align}
where $m_b=4.2\;{\rm GeV}$, $m_c=1.3\;{\rm GeV}$ and $m_\tau=1.8\;{\rm GeV}$.

The reheating temperature may be evaluated\footnote{We assume nonlinear effects
\cite{Dolgov:1989us,Traschen:1990sw,Kawai:2015lja} are negligible.
} by comparing the decay width with the Hubble expansion rate,
\begin{align}
	\Gamma_{\widetilde\phi}
	=H(T_{\rm rh})
	=\frac{T_{\rm rh}^2}{M_{\rm P}}\sqrt{\frac{\pi^2 g_*}{90}}.
\end{align}
Using the number of relativistic degrees of freedom from the Standard Model $g_*=106.75$, the reheating temperature for our benchmark parameter range \eqref{eqn:gXrange}, \eqref{eqn:mphirange}, \eqref{eqn:thetarange} is found to be
\begin{align}\label{eqn:Trh}
	83.3\;{\rm TeV}\leq T_{\rm rh}\leq 174\;{\rm TeV}.
\end{align}


The observed baryon asymmetry of the Universe is naturally explained by the scenario of baryogenesis via leptogenesis \cite{Fukugita:1986hr}.
In our model, the right-handed neutrinos are thermally created during reheating, and the baryon number can be generated by the sphaleron processes during the electroweak phase transition.
Although the temperature \eqref{eqn:Trh} is lower than $\sim 10^9$ GeV that is necessary in generic scenarios \cite{Davidson:2002qv}, it is sufficiently higher than the lower bound for successful leptogenesis\footnote{The reheating temperature necessary for successful resonant leptogenesis is $T_{\rm rh}\gtrsim$ a few TeV. See e.g. \cite{Arai:2012em}.} when the resonant enhancement \cite{Flanz:1996fb,Pilaftsis:1997jf,Pilaftsis:2003gt} takes place.
Since nearly degenerate right-handed neutrino mass is necessary for the resonance, our assumption of the degenerate Majorana Yukawa couplings is not only for the sake of simplicity but is indeed necessary for the successful baryogenesis scenario.

\subsubsection{Dark matter candidate}

\begin{table}[t]
\begin{tabular}{l|cccc}
  &~~~$SU(3)_c$&~~~$SU(2)_L$&~~~$U(1)_Y$~~~&~~~$U(1)_X$\\
  \hline\\
  $\zeta_{L,R}$&${\BF 1}$&${\BF 1}$&$0$&$Q$\\\\
\end{tabular}
\caption{Representations and charges of the Dirac dark matter fermion $\zeta=(\zeta_L,\zeta_R)$.
For stability the $U(1)_X$ charge $Q$ is assumed to be different from $\pm 1$, $\pm 3$.}
\label{tab:DMfermion}
\end{table}

Although it is known that one of the right-handed neutrinos in the $U(1)_X$-extended Standard Model can play the role of dark matter \cite{Okada:2010wd}, for the benchmark parameter values used above, the abundance of the dark matter right-handed neutrino created in thermal processes is too small to account for the dark matter abundance of the present Universe (for a review, see \cite{Okada:2018ktp}).
We thus have to look for a candidate of dark matter outside the particle contents of TABLE I.

The simplest candidate is a Dirac fermion, which is singlet under the Standard Model gauge group and has a generic $U(1)_X$ charge $Q$ such that $|Q|\neq 1,3$ (TABLE II).
Let us call it $\zeta=(\zeta_L, \zeta_R)$.
Unless $Q=\pm 1$ (in this case $\zeta$ has the same interaction as the right-handed neutrinos) or $Q=\pm 3$ (gauge invariant $\Phi^*\overline{N_R^c}\zeta_R$, $\Phi\overline{N_R}\zeta_L$ interactions can exist), $\zeta$ is protected by the $U(1)_X$ symmetry and thus is stable to be dark matter.
Adding the Dirac fermion does not spoil the anomaly cancellation of the $U(1)_X$-extended Standard Model.
The $\zeta$ field communicates with the Standard Model particles only through the $U(1)_X$ gauge interactions. 

The $\zeta$ dark matter can be produced thermally or non-thermally.
Let us consider the thermal production scenario first.
In this case, $\zeta$ is assumed to be in thermal equilibrium with the Standard Model particles through the $U(1)_X$ interactions $\zeta\overline\zeta\leftrightarrow Z'\leftrightarrow f_{\rm SM}\overline{f_{\rm SM}}$, and then freezes out \cite{FileviezPerez:2019jju}.
When the mass of $\zeta$ is half of $M_{Z'}$, the production $Z'\to\zeta\overline\zeta$ is enhanced by the resonance process for which the cross section is
\begin{align}\label{eqn:TDMcs}
  \langle\sigma v_{\rm rel}\rangle\simeq \frac{Q^2 g_X^2}{m_\zeta^2}.
\end{align}
In the freeze-out scenario, it is well known that the present dark matter abundance $\Omega_{\rm DM}h^2=0.12$ is obtained\footnote{
Only in this subsection $h$ denotes the normalized Hubble parameter.
} when the cross section is $\langle\sigma v_{\rm rel}\rangle\sim 1\;{\rm pb}$.
For our benchmark parameter value $M_{Z'}=6\;{\rm TeV}$, the dark matter mass is $m_\zeta=3\;{\rm TeV}$ and one can see from \eqref{eqn:TDMcs} that the dark matter abundance $\Omega_{\rm DM}h^2=0.12$ is reproduced if $Q\simeq 10$.

When $|Q|\lesssim{\C O}(1)$, the thermal dark matter scenario is not applicable as the $U(1)_X$ interaction is not strong enough to keep the Dirac fermion $\zeta$ in thermal equilibrium.
In this case, it is appropriate to consider non-thermal production of dark matter by the freeze-in mechanism.
After reheating, the dark matter $\zeta$ is produced from the Standard Model thermal plasma by the out-of-equilibrium processes,
\begin{align}\label{eqn:NTDMproc}
  f_{\rm SM}\overline{f_{\rm SM}}\to &\; Z'\to\zeta\overline\zeta,\crcr
  Z'Z'\to &\; \zeta\overline\zeta.
\end{align}
By numerically solving the Boltzmann equations for these processes, it is known \cite{Mohapatra:2019ysk} that the dark matter abundance 
$\Omega_{\rm DM}h^2=0.12$ is reproduced if
\begin{align}\label{eqn:NTDMcond}
  (Qg_X)^2 (x_Hg_X)^2+\frac{0.82}{1.2}(Qg_X)^4\simeq 8.2\times 10^{-24}.
\end{align}
When $|Q|\ll 1$, the first term of \eqref{eqn:NTDMcond} (the first line of \eqref{eqn:NTDMproc}) dominates. 
For our benchmark values $x_H={\C O}(10)$ and $g_X\simeq (\text{a few})\times 10^{-2}$, one can see that $|Q|\sim 10^{-9}$ gives the present dark matter abundance $\Omega_{\rm DM}h^2=0.12$.
The computation of \eqref{eqn:NTDMcond} assumes $M_{Z'}^2\ll m_\zeta^2$.
This condition is easily satisfied by, for example, choosing $m_\zeta=20\;{\rm TeV}$ for $M_{Z'}=6\;{\rm TeV}$.
Also, the reheating temperature of $T_{\rm rh}\sim 100\;{\rm TeV}$ is sufficient for producing the thermal plasma.


\section{Final remarks} 

We have discussed the particle physics implications of the cosmological data-fitting on the Coleman-Weinberg type effective potential, based on the recent MCMC analysis made in \cite{Rodrigues:2020dod}.
We focused on the particular inflationary scenario based on the $U(1)_X$-extended Standard Model, and analyzed the RG flow from the inflationary (Planck) scale.
By the RG analysis we identified the parameter region for which the $U(1)_X$ symmetry breaking due to the Coleman-Weinberg mechanism is operative. 
Moreover, we found the parameter region that is already excluded by the LHC Run-2, and the region that is covered by the future HL-LHC.

Data fitting by the cosmological MCMC combined with the RG analysis opens up a new direction of research in particle cosmology.
Although we focused on just one example of inflationary model here, a wide range of inflationary models based on gauge-extended Standard Models may be analyzed in a similar manner.
This approach is, in a sense, amalgamation of top-down (Planck scale physics) and bottom-up (collier physics) approaches.
The link between the two is provided by the RG flow.

\begin{acknowledgments}
This work was supported in part by the National Research Foundation of Korea Grant-in-Aid for Scientific Research 
NRF-2018R1D1A1B07051127
and the NRF-JSPS collaboration ``String Axion Cosmology" (S.K), by the United States Department of Energy Grants DE-SC0012447 (N.O.) and by the M. Hildred Blewett Fellowship of the American Physical Society, www.aps.org (S.O.).

\end{acknowledgments} 
 
\newpage
\appendix

\begin{widetext}
\section{1-loop renormalization group equations}

General discussions on renormalization group in the presence of multiple $U(1)$'s are found for example in \cite{delAguila:1988jz,Chankowski:2006jk}.
For the $U(1)_X$-extended Standard Model, with the matter content of Table I and the gauge coupling in the form of the covariant derivative \eqref{eqn:CovDeriv}, the 1-loop beta functions in the gauge sector are
($t\equiv \ln\mu$)
\begin{align}
  \frac{dg_3}{dt}=&\frac{1}{16\pi^2}\left(-7g_3^3\right),\\
  \frac{dg_2}{dt}=&\frac{1}{16\pi^2}\left(-\frac{19}{6}g_2^3\right),\\
  \frac{dg_1}{dt}=&\frac{1}{16\pi^2}\cdot\frac{41}{6}g_1^3,\\
  \frac{dg_X}{dt}=&\frac{1}{16\pi^2}\left(
  \frac{41}{6}g_X(\widetilde{g}_1+g_Xx_H)^2
  +\frac{32}{3}g_X^2(\widetilde{g}_1+g_X x_H)
  +12 g_X^3
\right),\\
  \frac{d\widetilde g_1}{dt}=&\frac{1}{16\pi^2}\left(
  \frac{41}{6}(\widetilde{g}_1+g_X x_H)
  (\widetilde g_1^2+2g_1^2+\widetilde{g}_1g_Xx_H)
   +\frac{32}{3}g_X(g_1^2+\widetilde{g}_1^2+\widetilde{g}_1 g_Xx_H)
   +12 g_X^2\widetilde{g}_1
\right).
\end{align}

The running of the top and Majorana Yukawa couplings is determined by
\begin{align}
  \frac{dy_t}{dt}=&\frac{y_t}{16\pi^2}\left(
  \frac{9}{2} y_t^2
  -8 g_3^2
  -\frac{9}{4}g_2^2
  -\frac{17}{12}g_1^2
  -\frac{17}{12}(\widetilde{g}_1+g_Xx_H)^2
  -\frac{2}{3}g_X^2
  -\frac{5}{3}(\widetilde{g}_1+g_Xx_H)g_X
\right),\\
  \frac{dy_M^i}{dt}=&\frac{y_M^i}{16\pi^2}\Big((y_M^i)^2+\half\Tr (y_M^i)^2-6g_X^2\Big).
\end{align}

The beta functions for the scalar self couplings are
\begin{align}
  \label{eqn:betaH}
  \beta_H\equiv &\frac{d\lambda_H}{dt}=\,\frac{1}{16\pi^2}\Big( 
  24\lambda_H^2+\widetilde\lambda^2-6y_t^4+12\lambda_H y_t^2
  -9 g_2^2\lambda_H-3g_1^2 \lambda_H
  -3\lambda_H(\widetilde g_1+g_Xx_H)^2\crcr
  &\qquad\qquad
  +\frac{9}{8}g_2^4+\frac{3}{8}(\widetilde{g}_1+g_Xx_H)^4
  +\frac{3}{8}g_1^4
  +\frac{3}{4}g_1^2 g_2^2
  +\frac{3}{4}(g_1^2+g_2^2)(\widetilde{g}_1+g_Xx_H)^2
\Big),\\
  \label{eqn:betaPhi}
  \beta_\Phi\equiv &\frac{d\lambda_\Phi}{dt}=\,\frac{1}{16\pi^2} \Big(20\lambda_\Phi^2+2\widetilde\lambda^2-\Tr (y_M^i)^4
  +96 g_X^4+2\lambda_\Phi\Tr (y_M^i)^2-48\lambda_\Phi g_X^2\Big),\\
  \label{eqn:betamix}  
  \beta_{\rm mix}\equiv &\frac{d\widetilde\lambda}{dt}=\,\frac{\widetilde\lambda}{16\pi^2}\Big(
  12\lambda_H+8 \lambda_\Phi+4\widetilde\lambda+6 y_t^2-\frac{9}{2}g_2^2
-\frac 32 g_1^2
-\frac 32(\widetilde g_1+g_Xx_H)^2\crcr
&\qquad\qquad
+\Tr (y_M^i)^2-24 g_X^2
+12g_X^2(\widetilde g_1+g_Xx_H)^2\widetilde\lambda^{-1}
  \Big).
\end{align}


\end{widetext}

\providecommand{\href}[2]{#2} 
%
 

\end{document}